\documentstyle[12pt]{article}
\input tcilatex

\begin{document}

\begin{center}
{\bf STOCHASTIC EVOLUTION OF COSMOLOGICAL PARAMETERS IN THE EARLY UNIVERSE}\\
\vspace{2cm} {\bf C. Sivakumar$^{\dag}$, Moncy V. John$^{\ddag}$ and K. Babu
Joseph}\\ Department of Physics, Cochin University of Science and Technology 
\\ Kochi 682022, India\\

\vspace{1.5cm} {\bf Short title :} Stochastic Approach To Early Universe\\
\end{center}

\vspace{2cm} \noindent {\bf Abstract.} We develop a stochastic formulation
of cosmology in the early universe, after considering the scatter in the
redshift-apparent magnitude diagram in the early epochs as an observational
evidence for the non-deterministic evolution of early universe. We consider
the stochastic evolution of density parameter in the early universe after
the inflationary phase qualitatively, under the assumption of fluctuating $w$
factor in the equation of state, in the Fokker-Planck formalism. Since the
scale factor for the universe depends on the energy density, from the
coupled Friedmann equations we calculated the two variable probability
distribution function assuming a flat space geometry.\\

\vspace{1cm} \noindent {\bf Key words.} Cosmology; stochastic equations;
Fokker-Planck equation.\\

\vspace{1cm} \noindent {\bf PACS Nos 98.80; 02.50 }\\

\vspace{.5cm} \noindent {\dag} e-mail: sivakumarc @ cusat. ac. in\\ {\ddag}
Permanent Address: Department of Physics, St. Thomas College,\\ 
\hspace{1.5cm} Kozhencherri, Kerala, India 689641 e-mail: moncy @ stthom.
ernet. in\\

\newpage
\baselineskip 28pt \noindent{\large {\bf 1. Introduction}}

\vspace{.5cm}

\noindent The simplest model of the universe \cite{Peebles}, called the
Friedmann model or hot big bang model, is based on the assumption that the
matter distribution in the universe is homogeneous and isotropic on very
large scales (cosmological principle). This assumption simplifies the
problem of solving Einstein's equations. The assumption of homogeneity leads
to the prediction of deterministic Hubble's law and one expects a scatter
free redshift-magnitude diagram for galaxies and other extra galactic
objects. For nearer galaxies, Hubble's law is justified by a relatively
scatter-free redshift-magnitude relation. This corresponds to a
deterministic evolution of the universe. However, for high redshift quasars
(in the Hewitt-Burbidge catalogue [2,3]) and for the latest Type Ia
supernovae data [4,5], the redshift-magnitude diagram is a scatter diagram,
i.e., they are in contrast with the deterministic Hubble's law. If Hubble's
law is valid for all extra-galactic objects, then the evolution of the
universe (or expansion rate) is non-deterministic in the early epoch, as
indicated by the scatter diagram. We observe that the scatter increases as
we probe into more and more distant epochs. The small scatter for galaxies
at low redshifts is explained as due to peculiar velocities and the
conventional explanation for the peculiar velocities is that they are
induced by the observed density perturbations. This may be adequate to
account for the observed peculiar velocities of objects, in the range of 100
km/s - 400 km/s. Thus this may lead to a very small scatter at low
redshifts. But since the amplitude of density perturbations in the early
universe was very low, the large scatter at very high redshifts remains
unexplained and it is desirable to look for some alternative mechanism. Some
authors speculate [3,6] that the quasars are not at their cosmological
distances and have proposed some non-cosmological contributions to $z$ as a
possible explanation for the scatter in the quasar data. Since the distance
measurements are extremely difficult for quasars, due to the difficulty in
identifying standard candles, the scatter may be caused by the variation of
intrinsic luminosities of quasars of same $z$. However, the data is more
accurate for supernovae and the scatter in its redshift-magnitude diagram is
not due to either peculiar velocities or variation of intrinsic
luminosities. In \cite{Siva}, we proposed that a fluctuating equation of
state or a fluctuating mean $w$ factor in the equation of state led to a
non-deterministic or stochastic Hubble parameter and argued that such a
fluctuating expansion rate in the early universe might have led to a
randomness in the recession velocities of objects, in addition to peculiar
velocities and will produce a scatter in the redshift-magnitude diagram in
those epochs. Here we develop a more general description of the stochastic
dynamics of the early universe, and discuss the non-deterministic character
of total density of the universe, as well as the scale factor for the
universe. Such a stochastic approach is necessary, when the mean $w$ factor
in the equation of state of the cosmic fluid is a fluctuating quantity.

\vspace{.5cm} \noindent{\large {\bf 2. Observational issues and assumptions}}

\vspace{.5cm}

\noindent In this section, we state the conditions under which a stochastic
equation of state (or a fluctuating mean $w$ factor) emerges. In standard
cosmology, the cosmological fluid is in fact not unicomponent, instead
matter and radiation (with equation of state $p_m=0$ and $p_r=\left( \frac
13\right) \rho _r$, respectively) in disequilibrium coexist in many
`elementary subvolumes' of the universe \cite{Pavon}. Some recent
measurements on the age of the universe, Hubble parameter, deceleration
parameter, gravitational lensing etc., point to the need of extending the
standard model by including some new energy density (missing energy) in the
present universe, in addition to the usual relativistic/non-relativistic
energy density. Recent observations using Type Ia supernovae [4,5] as
standard candles seem to indicate that the universe may be accelerating,
driven by a positive cosmological constant (or a vacuum energy density $\rho
_v$) with the equation of state $p_v =-\rho _v$. Some authors [9,10]
introduced Quintessence or Q-component (for exa: scalar fields rolling down
a potential with negative pressure or cosmic strings etc.) with equation of
state $p_Q=w_Q $ $\rho _Q$ , $-1<w_Q\leq 0,$ in addition to matter density.
The case for a positive cosmological constant has been considered very often
in the literature [11-23]. All these models favour a flat universe with $%
\Omega=\Omega _m+\Omega _v+\cdot \cdot \cdot =1,$ where $\Omega $ is the
ratio of energy density to critical density. Thus our universe may be
approximated by a perfect fluid having many components, each with equation
of state $p_i=w_i\rho _i$, $-1\leq w_i\leq +1$ , $i=1,2\cdot \cdot \cdot $.
If we denote the total energy density due to all such components as $\rho $,
then

\begin{equation}
\label{eq:1}\rho =\rho _m+\rho _r+\rho _v+\cdot \cdot \cdot , 
\end{equation}

\noindent where $\rho _m,$ $\rho _r,$ $\rho _v$ etc. are the average
densities of matter, radiation, vacuum energy etc. In a similar way the
total pressure $p$ can be written as 
\begin{equation}
\label{eq:2}p=p_m+p_r+p_v+\cdot \cdot \cdot 
\end{equation}

\noindent In general, $\rho =\sum \rho _i$, where $\rho _i$'s represent
energy densities of various components. From the energy-momentum
conservation law (here it is assumed that only the total energy density is
conserved), we have

\begin{equation}
\label{eq:3}\dot \rho =-3\frac{\dot a}a(\rho +p)=-3\frac{\dot a}a\rho (1+w), 
\end{equation}
where $p=\sum p_i$ is the total pressure, and the ratio $w=p/\rho $ should
lie between -1 and +1. Splitting $\rho $ and $p$ into individual components,
the above equation becomes

\begin{equation}
\label{eq:4}\dot \rho _1+\dot \rho _2+\dot \rho _3+\cdot \cdot \cdot \cdot
=-3 \frac{\dot a}a[\rho _1(1+w_1)+\rho _2(1+w_2)+\rho _3(1+w_3)+\cdot \cdot
]. 
\end{equation}
From Eqs. (\ref{eq:3}) and (\ref{eq:4}), we get

\begin{equation}
\label{eq:5}w(t)=-\frac{\left[ \dot \rho _1+\dot \rho _2+\cdot \cdot \cdot
\right] }{3(\dot a/a)\rho }-1=\frac{\left[ \rho _1(1+w_1)+\rho
_2(1+w_2)+\cdot \cdot \cdot \cdot \right] }\rho -1. 
\end{equation}
\noindent Here we assumed that the total energy density $\rho $ is conserved
and not the parts corresponding to $\rho _1,\rho _2\cdot \cdot \cdot $
separately, and hence there can be creation of one component at the expense
of other components and since $\dot a/a$ and $\rho $ are large in the
earlier epochs, at least some of the $\dot {\rho _i}$'s will be
significantly large in this period. Since recent observations indicate the
existence of vacuum energy even in the present universe, one can expect that
particle creation continued for a fairly long period in the early universe 
\cite{Carvalho}. If we have a many component fluid, then the Einstein
equations, along with the equations of state of individual components, are
insufficient to determine the creation rates of each component. In \cite
{Wein}, Weinberg discusses the consequences of the presence of a
cosmological constant (vacuum energy) in the energy density. He discusses
some phenominological proposals made by some authors, of the energy transfer
between vacuum and matter or vacuum and radiation, in such a way that either 
$\rho _v/\rho _m$ or $\rho _v/\rho _r$ remains constant, respectively. He
also considers the possibility of creation of radiation from vacuum energy,
keeping $\rho _v/\rho _m$ fixed. However, in a general case as in Eq. (\ref
{eq:5}) above, one cannot expect any kind of creation in the universe to be
a smooth process, since they can be sporadic events occurring in different
locations and times, like those occurring in galactic nuclei. Here, as in
the case of other stochastic processes like Brownian motion, a complete
solution of the macroscopic system (universe) would consist in solving all
the microscopic equations describing the creation processes, but such a
rigorous derivation will be very complicated or even impossible. In this
context, a stochastic approach is more reasonable, in which we consider the
creation rates to be fluctuating, leading to fluctuations in the ratios $%
\rho _i/\rho .$ As it clear from Eq. (\ref{eq:5}), this in turn, will lead
to a fluctuating $w$ factor, which is the key assumption made in this paper.
Such a fluctuating quantity modifies the dynamics of the early universe,
where the evolution of the cosmological parameters, like the total energy
density $(\rho )$ and scale factor for the universe ($a)$ becomes stochastic
or non-deterministic. The fluctuations in the ratio $\rho _i/\rho$, that we
are taking in to account here are classical, i.e., our stochastic model is a
modification to the classical Friedmann model of the early universe, when
fluctuations are significant. In \cite{Fang}, Fang {\sl et al} discuss a
stochastic approach to early universe (before recombination epoch), when
cosmic fluid consisting of primeval plasma and radiation, is not perfect,
but have dissipations due to differences in the adiabatic cooling rates of
the components of the fluid and the possible energy transfer between them.
Physically motivated interaction models are also proposed in the literature
[26,27], which lead to energy transfer between various components. However,
once we probe into still earlier epochs (stages of inflation etc.), quantum
fluctuations become very important. Many authors discuss the need for a
stochastic approach to inflation [28,29,30], when the quantum fluctuations
of the scalar field are significant, and try to get a probability
distribution function for the scalar field after solving the quantum
Langevin equation (or FPE) describing the evolution of the scalar field.
However we adopt the stochastic approach in the classical regime, where
fluctuations in the creation rates and also in the possible energy transfer
between different components of the cosmic fluid lead to a stochastic
equation of state. This causes a non-deterministic (stochastic) expansion
rate for the universe, described by a set of stochastic differential
equations, instead of the deterministic Friedmann equations and we evaluate
the probability distribution function of the cosmological parameters. In
section 3 we discuss the stochastic character of the density parameter of
the universe on the basis of Fokker-Planck formalism \cite{Fokk} . In
section 4 we analyse the effect of such phenomena on the expansion factor
for the universe and a two variable distribution function is derived.

\vspace{.5cm}

\noindent{\large {\bf 3. Stochastic evolution of density parameter}}

\vspace{.5cm}

\noindent Suppose the universe is approximated by a many component fluid in
the early epochs, with a fluctuating $w$ term in the equation of state. Now
we write the evolution equation for the total density in the early universe
(assuming that, total energy density is conserved), immediately after
inflation, when curvature factor appearing in the field equation is
negligible, so that the background is approximately flat.

\begin{equation}
\label{eq:6}\dot \rho =-3\frac{\dot a}a[1+w(t)]\rho . 
\end{equation}

\noindent Overdots denote time derivatives. Using Friedmann equations we have

\begin{equation}
\label{eq:7}\dot \rho =-\sqrt{24\pi G}[1+w(t)]\rho ^{3/2}. 
\end{equation}

\noindent Above equation is a stochastic differential equation of the
Langevin type. Since $w$ is a fluctuating `force' term, $\rho $ is a
stochastic variable or its evolution is non-deterministic. The random
behaviour of $\rho $ in the early universe is due to fluctuations in the
factor $w$ alone. If fluctuations are zero we are back to the deterministic
standard model. We apply stochastic methods \cite{Fokk} for the analysis of
the above equation and the probability distribution function is calculated
using Fokker-Planck formalism. By making use of the transformation 
\begin{equation}
\label{eq:8}\sigma =\frac 1{(6\pi G\rho )^{1/2}}, 
\end{equation}

\noindent eq. (\ref{eq:7}) gets modified into 
\begin{equation}
\label{eq:9}\dot \sigma =1+w(t), 
\end{equation}

\noindent which is a non-deterministic, stochastic first order differential
equation (Here $\sigma \propto t$ for a pure deterministic case). To solve
eq. (\ref{eq:9}) we use certain simplifying assumptions that the fluctuating
factor $w$ is Gaussian $\delta $-correlated, with mean zero. Though these
assumptions are taken for the sake of simplicity, we expect that they are
reasonable when compared to the time scales involved.

\noindent If we have a general Langevin type equation of the form 
\begin{equation}
\label{eq:10}\dot y=h(y,t)+g(y,t)\Gamma (t), 
\end{equation}

\noindent where $\Gamma (t)$ is a fluctuating quantity with zero mean and
Gaussian $\delta $-correlated, then the corresponding Fokker-Planck equation
(FPE) describing the time evolution of the probability distribution function 
$W(y,t)$ can be written as 
\begin{equation}
\label{eq:11}\frac{\partial W(y,t)}{\partial t}=\sum_{n=1}^\infty \left(
-\frac \partial {\partial y}\right) ^nD^{(n)}(y)W(y,t) , 
\end{equation}

\noindent where $D^{(n)}(y)$ are the Kramers Moyal expansion coefficients
given by 
\begin{equation}
\label{eq:12}D^{(n)}(y,t)=\frac 1{n!}\left[ \lim _{\tau \rightarrow 0}\frac
1\tau \left\langle [y(t+\tau )-x]^n\right\rangle \right] _{y(t)=x} . 
\end{equation}

\noindent Here $y(t+\tau )$ $(\tau >0)$ is a solution of eq. (\ref{eq:10})
which at time $t$ has the sharp value $y(t)=x.$ Under the assumption of $%
\delta $ -correlation and zero mean of $\Gamma (t),$ all coefficients
vanishes for $n\geq 3$ and retain only the coefficients $D^{(1)}$ and $%
D^{(2)},$ called drift and diffusion coefficients respectively. Following
the standard procedure \cite{Fokk}, eq. (\ref{eq:9}) leads to the FPE 
\begin{equation}
\label{eq:13}\frac{\partial W(\sigma ,t)}{\partial t}=-\frac{\partial W}{
\partial \sigma }+D\frac{\partial ^2W}{\partial \sigma ^2} . 
\end{equation}

\noindent Here we used drift coefficient $D^{(1)}=1$ and diffusion
coefficient $D^{(2)}=D$ ($D$ is a constant with dimension of time, which is
introduced for the purpose of generality). In order to obtain non-stationary
solutions of eq. (\ref{eq:13}) we use a separation ansatz for $W(\sigma ,t)$ 
\begin{equation}
\label{eq:14}W(\sigma ,t)=\phi (\sigma )e^{-\lambda t}. 
\end{equation}

\noindent Substituting this into eq. (\ref{eq:13}) and solving for $\phi
(\sigma )$ we get

\begin{equation}
\label{eq:15}\phi (\sigma )=A\exp [\frac \sigma {2D}+ik\sigma ], 
\end{equation}

\noindent where 
\begin{equation}
\label{eq:16}k=\pm \sqrt{\frac \lambda D-\frac 1{4D^2}}. 
\end{equation}

\noindent Thus we see that for $\lambda <1/4D,$ $k^2$ is negative and the
solution is exponentially diverging, which is not a physically reasonable
solution. Hence we conclude that $\lambda \geq 1/4D,$ so that $k$ is real.
We write the most general solution as 
\begin{equation}
\label{eq:17}W\left( \sigma ,t\right) =\sum_nc_n\phi _n\left( \sigma \right)
e^{-\lambda _nt}, 
\end{equation}

\noindent where $c_n$ can be real or complex but $W\left( \sigma ,t\right) $
is always real. For a continuous parameter $k,$ from eqs. (\ref{eq:14}) and
( \ref{eq:15}) the general solution or the distribution function is given by 
\begin{equation}
\label{eq:18}W\left( \sigma ,t\right) =A\int_{-\infty }^{+\infty }\exp
[\frac \sigma {2D}+ik\sigma -k^2Dt-\frac t{4D}]dk. 
\end{equation}

\noindent We choose $A=1/2\pi $ for normalization purpose. On evaluating the
integral we find the distribution function 
\begin{equation}
\label{eq:19}W\left( \sigma ,t\right) =\frac 1{\sqrt{4\pi Dt}}\exp \left[ - 
\frac{\left( \sigma -t\right) ^2}{4Dt}\right] , 
\end{equation}

\noindent which is in Gaussian form. The expectation value of the stochastic
variable $\sigma $ is $\left\langle \sigma \right\rangle =t$ and corresponds
to the deterministic solution of eq. (\ref{eq:9}). The width of the Gaussian
is found from the variance, $v=\left\langle \left( \sigma -\left\langle
\sigma \right\rangle \right) ^2\right\rangle $ = $2Dt.$ Once $W(\sigma ,t)$
is known it is straight forward to write the distribution function $W(\rho
,t)$ as 
\begin{equation}
\label{eq:20}W\left( \rho ,t\right) =\frac 1{\sqrt{96\pi ^2DG\rho ^3t}}\exp
\left[ -\frac{\left( 1-t\sqrt{6\pi G\rho }\right) ^2}{24\pi G D\rho t}
\right] . 
\end{equation}

\noindent We can also find the transition probability for the stochastic
variable to change from an initial state $\left( \sigma ^{\prime },t^{\prime
}\right) $ to a final state $\left( \sigma ,t\right) $ as 
\begin{equation}
\label{eq:21}P\left( \sigma ,t\mid \sigma ^{\prime },t^{\prime }\right)
=\frac 1{\sqrt{4\pi D\left( t-t^{\prime }\right) }}\exp \left[ -\frac{\left[
\left( \sigma -\sigma ^{\prime }\right) -\left( t-t^{\prime }\right) \right]
^2}{ 4D\left( t-t^{\prime }\right) }\right] , 
\end{equation}

\noindent with the initial value 
\begin{equation}
\label{eq:22}P\left( \sigma ,t\mid \sigma ^{\prime },t\right) =\delta \left(
\sigma -\sigma ^{\prime }\right) , 
\end{equation}

\noindent indicating Markovian nature of the random variable $\sigma .$ In
terms of $\rho $ eq. (\ref{eq:21}) becomes

\begin{equation}
\label{eq:23}P\left( \rho ,t\mid \rho ^{\prime },t^{\prime }\right) =\frac 1{%
\sqrt{4\pi D\left( t-t^{\prime }\right) }}\exp \left[ -\frac{\left[ \left( 
\sqrt{\rho ^{\prime }}-\sqrt{\rho }\right) -\sqrt{6\pi G\rho \rho ^{\prime }}
\left( t-t^{\prime }\right) \right] ^2}{24\pi G\rho \rho ^{\prime }D\left(
t-t^{\prime }\right) }\right] 
\end{equation}

\noindent This represent the probability for the energy density to change
from an initial value $\rho $ to a final value $\rho ^{\prime }$ during a
time interval $\left( t-t^{\prime }\right) $ in the early epochs. This
characterises the stochastic behaviour of density evolution in the early
universe.

\vspace{.5cm}

\noindent{\large {\bf 4. Scale factor as a stochastic variable}}

\vspace{.5cm}

\noindent Under the assumption that the factor $w$ is fluctuating during the
early epochs, the evolution of the scale factor also becomes
non-deterministic, since the time evolution of $a\left( t\right) $ is
determined by the total density. So we have a system of coupled stochastic
differential equations derived from Friedmann equations

\begin{equation}
\label{eq:24}\dot a=\sqrt{\frac{8\pi G}3}\rho , 
\end{equation}

\noindent and

\begin{equation}
\label{eq:25}\dot \rho =-\sqrt{24\pi G}\left[ 1+w(t)\right] \rho ^{3/2}. 
\end{equation}

\noindent Here we are considering the dynamics of the universe immediately
after inflation, so that the background can be treated approximately flat.
With the transformation defined in eq. (\ref{eq:8}), the above system of
equations reduce to 
\begin{equation}
\label{eq:26}\dot a=\frac{2a}{3\sigma }, 
\end{equation}

\noindent and 
\begin{equation}
\label{eq:27}\dot \sigma =1+w(t). 
\end{equation}

\noindent Following the standard procedure \cite{Fokk} we have the drift
coefficients $D_a^{\left( 1\right) }=\frac{2a}{3\sigma },$ $D_\sigma
^{\left( 1\right) }=1$ and the diffusion coefficient $D_{\sigma \sigma
}^{\left( 2\right) }=D$ is assumed to be a constant. It shall be noted that
this diffusion term arises due to fluctuations in $w$ alone. The two
variable FPE for the distribution function $W\left( a,\sigma ,t\right) $ can
be written as 
\begin{equation}
\label{eq:28}\frac{\partial W}{\partial t}=-\frac 2{3\sigma }\left[ W+a\frac{
\partial W}{\partial a}\right] -\frac{\partial W}{\partial \sigma }+D\frac{
\partial ^2W}{\partial \sigma ^2} . 
\end{equation}

\noindent We can solve the FPE by first assuming the ansatz 
\begin{equation}
\label{eq:29}W\left( a,\sigma ,t\right) =U(a)V(\sigma )e^{-\lambda t} , 
\end{equation}

\noindent and substituting into eq. (\ref{eq:28}). We obtain 
\begin{equation}
\label{eq:30}\frac \sigma VD\frac{d^2V}{d\sigma ^2}-\frac \sigma V\frac{dV}{
d\sigma }+\lambda \sigma =\frac 23\left[ \frac aU\frac{dU}{da}+1\right] . 
\end{equation}

\noindent Each side in this equation can be equated to a constant $m$. When $%
m=0$

\begin{equation}
\label{eq:31}U(a)\propto \frac 1a, 
\end{equation}

\noindent and 
\begin{equation}
\label{eq:32}V\left( \sigma \right) \propto \exp [\frac \sigma {2D}+ik\sigma
] , 
\end{equation}

\noindent with $k$ given by eq. (\ref{eq:16}). A physically reasonable
solution exists for $\lambda \geq 1/4D,$ which is 
\begin{equation}
\label{eq:33}W\left( a,\sigma ,t\right) =\frac Ba\exp \left[ \frac \sigma
{2D}+ik\sigma -\lambda t\right] . 
\end{equation}

\noindent Here $B$ is a normalization constant, chosen to be $1/2\pi .$ One
point to be noted is the most general solution to eq. (\ref{eq:30}) when $%
m\neq 0$, is a series solution owing to the singularity at $\sigma =0.$ One
can find a limiting solution as $\sigma \rightarrow 0,$ in the following
form 
\begin{equation}
\label{eq:34}W\left( a,\sigma \right) \longrightarrow a^{(\frac 23m-1)}\frac{
\exp \left( \frac \sigma {2D}\right) }{\left( m/D\right) }\sum_{n=1}^\infty 
\frac{\left( \frac mD\sigma \right) ^n}{n!\left( n-1\right) !} . 
\end{equation}

\noindent However we will get a real general solution in a compact form
after integrating eq. (\ref{eq:33}) in the range $-\infty <k<+\infty $ 
\begin{equation}
\label{eq:35}W\left( a,\sigma ,t\right) =\frac 1{\sqrt{4\pi Dt}}\left(
a\right) ^{-1}\exp \left[ -\frac{\left( \sigma -t\right) ^2}{4Dt}\right] . 
\end{equation}

\noindent In terms of $\rho ,$ it becomes

\begin{equation}
\label{eq:36}W\left( a,\rho ,t\right) =\frac 1{\sqrt{96\pi ^2DG\rho ^3a^2t}
}\exp \left[ -\frac{\left( 1-t\sqrt{6\pi G\rho }\right) ^2}{24\pi GD\rho t}
\right] . 
\end{equation}

\noindent The two variable probability distribution function is Gaussian in $%
\sigma ,$ and diverges as $a\rightarrow 0,$ where classical approach fails
and quantum theory takes over. Now we write the expression for the
transition probability 
\begin{equation}
\label{eq:37}P\left( a,\sigma ,t\mid a^{\prime },\sigma ^{\prime },t^{\prime
}\right) =\frac{\left( aa^{\prime }\right) ^{-1}}{\sqrt{4\pi D\left(
t-t^{\prime }\right) }}\exp \left[ -\frac{\left[ \left( \sigma -\sigma
^{\prime }\right) -\left( t-t^{\prime }\right) \right] ^2}{4D\left(
t-t^{\prime }\right) }\right] . 
\end{equation}

\noindent In terms of $\rho $ and $a$ it becomes 
\begin{equation}
\label{eq:38}P\left( a,\rho ,t\mid a^{\prime },\rho ^{\prime },t^{\prime
}\right) =\left( aa^{\prime }\right) ^{-1}\frac{\exp \left[ -\frac{\left[
\left( \sqrt{\rho ^{\prime }}-\sqrt{\rho }\right) -\sqrt{6\pi G\rho \rho
^{\prime }}\left( t-t^{\prime }\right) \right] ^2}{24\pi G\rho \rho ^{\prime
}D\left( t-t^{\prime }\right) }\right] }{\sqrt{4\pi D\left( t-t^{\prime
}\right) }} , 
\end{equation}

\noindent which represents the transition probability for the variables to
change from the state ($a^{\prime },\rho ^{\prime })$ to $\left( a,\rho
\right) .$ Thus the scale factor $a$ together with the density $\rho $
evolves in a non-deterministic way, which in turn strongly influence the
formation of large scale structure in the universe, since the evolution of
the density perturbations are also depend on $w.$ In all these cases we get
Gaussian distributions, which is sharply peaked initially but spread out
with time.

\vspace{.5cm}

\noindent{\large {\bf 5. Conclusion}}

\vspace{.5cm}

\noindent In the preceding sections we have described a stochastic approach
to cosmology as a modification to the deterministic evolution of the
universe in the standard model. In section 2, we have shown that
fluctuations in the creation rates are physical processes which can lead to
a stochastic equation of state. A fluctuating $w$ factor, in turn, will lead
to fluctuations in the time - evolution of the energy density of the
universe, as well as in the expansion factor for the universe. Thus both
parameters become stochastic quantities, instead of remaining deterministic
variables. A fluctuating $w$ factor will also lead to fluctuations in the
time-evolution of the Hubble parameter; i.e., the expansion rate of the
universe becomes a stochastic quantity, instead of being a deterministic
variable. We predict that such a stochastic expansion rate might have led to
a randomness in the recession velocities of objects and will produce a
scatter in the Hubble diagram. Using the redshift-magnitude data $\mu _0$
for the Type Ia supernovae of Perlmutter {\sl et al} \cite{Perl99}, which
corresponds to their Fit C, and the technique used in Riess {\sl et al} \cite
{Riess}, we have computed the probability distribution function $p(H_0|\mu
_0)$ for a flat universe and compared it with the theoretical distribution 
\cite{Siva}, showing that both curves agree very well with $D=3.77\times
10^{13}$ s and a half width of $0.011$. A similar estimate is possible also
with the distribution function of the density parameter, provided we are
given a known data for the density parameter. Also if we have some explicit
examples of models where a stochastic $w$ emerges, the predicted value of $D$
may be compared with our estimation, but in this paper, we have not made any
attempts in this regard. Thus, if $w$ is a fluctuating quantity, the
evolution of the early universe becomes stochastic or non-deterministic (the
scatter in the Hubble diagram indicates this), and the dynamical equations
of those epochs are Langevin type equations, where one can evaluate the
probability distribution functions of the variables. In \cite{Berera},
Berera and Fang describe dynamically how, the stochastic fluctuations
arising from various dissipations, generate seeds of density perturbations
in the early universe, apart from the quantum fluctuations of the standard
inflationary model. Here we attempted to make a stochastic approach to the
early universe due to fluctuations in the mean equation of state, which is a
classical phenomenon. Since the stochastic equation of state, lead to
fluctuations in the time - evolution of the total density of the universe,
the density contrast will also be fluctuating. We propose to undertake this
study in detail in a future publication.

To conclude, we note that the stochastic approach presented above is a
modification to the standard model, when fluctuations are present. We have
formulated a stochastic model and developed a set of non-deterministic,
Langevin equations for the cosmological parameters in those epochs, under
the assumption that the factor $w(t)$ is a fluctuating quantity, when the
universe is approximated by a many component fluid. It is expected that
diffusion coefficient is a crucial factor in the evolution of the universe,
especially in the early phase, where it influences the time - evolution of
density parameter and scale factor for the universe. As fluctuations die out
with time ($D\rightarrow 0),$ the evolution becomes deterministic.

\vspace{.5cm}\noindent{\bf Acknowledgements}

\vspace{.5cm}

\noindent The authors thank the referee for valuable suggestions. One of us
(CS) thanks CSIR, New Delhi for the award of a Research Fellowship.

\end{document}